\documentclass[letter]{aa}  

\usepackage{graphicx}
\usepackage{txfonts}
\usepackage{lipsum}
\usepackage{subcaption}         % necessary for continued figures, example in section 3
\usepackage{lscape}             % to rotate a single page table, example in appendix.
\usepackage{placeins}           % useful with \FloatBarrier, to keep 
\usepackage{makecell}

\makeatletter
\let\do@linenumbers\relax
\let\linenumbers\relax
\makeatother

\begin{document}

   \title{Temporal Evolution of the Third Interstellar Comet 3I/ATLAS: Spin, Color, Spectra and Dust Activity\thanks{based on observations made with the Southern African Large Telescope (SALT)}}

   \subtitle{}

%%%%%%%%%%%%%%%%%%%%%%%%%%%%%%%%%%%%%%%%
% Please do not include ORCIDs next to author names.
% Only ORCIDs authenticated by individual authors in EDP Sciences editorial system will be taken into account.
% ORCIDs included here will be removed.
%%%%%%%%%%%%%%%%%%%%%%%%%%%%%%%%%%%%%%%%

   \author{T. Santana-Ros\inst{1,2}
   \and
   O. Ivanova\inst{3} %(Oleksandra is key to this paper, TK)
   \and
   S. Mykhailova\inst{4}  %(a key person for SALT spectrum, TK)
   \and
   N. Erasmus\inst{5,6} %(Nicolas started it all, observed at Lesedi,TK)
    \and
    K. Kamiński\inst{4} %(valuable help with data reduction, TK)
    \and
   D. Oszkiewicz\inst{4} %(a key person for NOT photometry, TK)
   \and
   T. Kwiatkowski\inst{4} 
   \and
   M. Hus\'arik\inst{3} %(valuable help with data reduction from SP, OI) ??
   \and
   T. S. Ngwane\inst{6,5}  %(provided het time at Lesedi, TK)
   \and
   A. Penttil\"{a}\inst{7} %(provided his time at NOT, TK)
        }

   \institute{
   Departamento de F\'{\i}sica, Ingenier\'{\i}a de Sistemas y Teor\'{\i}a de la Se\~{n}al, Universidad de Alicante, Carr. San Vicente del Raspeig, s/n, 03690 San Vicente del Raspeig, Alicante, Spain
   \and
   Institut de Ci\`encies del Cosmos (ICCUB), Universitat de Barcelona (UB), c. Mart\'i Franqu\`es, 1, 08028 Barcelona, Catalonia, Spain\\ 
   \email{tsantanaros@icc.ub.edu}
   \and
   Astronomical Institute of the Slovak Academy of Sciences, 059 60 Tatransk\'{a} Lomnica, Slovak Republic
    \and
    Astronomical Observatory Institute, Faculty of Physics and Astronomy, Adam Mickiewicz University, Słoneczna 36, 60-286 Poznań, Poland 
   \and
   South African Astronomical Observatory, 1 Observatory Rd, Observatory, Cape Town, 7925, South Africa
   \and
   Department of Physics, Stellenbosch University, Stellenbosch, 7602, South Africa.
    \and
    Department of Physics, PO Box 64, FI-00014 University of Helsinki, Finland
      }

   \date{Received 25 July 2025}

% \abstract{}{}{}{}{}
% 5 {} token are mandatory
 
  \abstract
{}
{We aim to characterize the physical and activity properties of the interstellar comet 3I/ATLAS through spectroscopic and photometric observations during the first month after its discovery.}
{We performed time-series photometry and long-slit spectroscopy between 2 and 29 July 2025 using multiple ground-based telescopes. Photometric data were calibrated against field stars from the ATLAS and APASS catalogs, and Fourier analysis was applied to derive the comet's rotational period. Spectral data were obtained using SALT and Nordic Optical Telescope.}
{We report a spin period of $16.16 \pm 0.01$~h with a lightcurve amplitude of approximately 0.3~mag. The comet exhibits increasing dust activity and reddening colors during the observation period, with no visible tail detected, likely due to viewing geometry and low dust production. Dust mass loss rates are estimated between 0.3 and 4.2~kg~s$^{-1}$, consistent with weakly active distant comets. Spectral colors are similar to those of outer Solar System comets and differ from previously reported values for 3I/ATLAS.}
% conclusions heading (optional)
{The morphological and photometric properties of 3I/ATLAS are consistent with a weakly active comet of outer Solar System origin, despite its interstellar provenance. Continued monitoring around perihelion is necessary to track changes in activity, color, which will provide insights into the evolution of interstellar materials under solar radiation.}


   \maketitle

%%%%%%%%%%%%%%%%%%%%%%%%%%%%%%%%%%%%%%%%%%%%%%%%%%%%%%%%%%%%%%
\section{Introduction}

Interstellar objects (ISOs) are minor bodies on hyperbolic, Sun-unbound trajectories, with excess speed at infinity confirming their extrasolar origin \citep{Meech17}. Each ISO provides a physical sample of extrasolar planetesimals. Studying them is compelling: their compositions, activity, and colors probe solid-body formation in other planetary systems, and the first two ISOs—1I/‘Oumuamua, an apparently inactive, extremely flattened body, and 2I/Borisov, a dust-rich comet—illustrate the diversity of extrasolar debris reservoirs \citep{Micheli18,Jewitt19}. Their trajectories reveal how Galactic tides, stellar flybys, and giant-planet scattering shape Oort clouds \citep{Seligman23Review}, while unusual orbits offer rare tests of sublimation and space weathering. Rapid-response ISO campaigns also refine strategies for anticipated Rubin Observatory discoveries \citep{Fitzsimmons23}.

The third confirmed ISO, comet 3I/ATLAS (C/2025 N1), was discovered on 1 July 2025 by the ATLAS 0.5~m telescope at Río Hurtado, Chile, and announced as interstellar the next day \citep{MPEC2025N12}. Archival ZTF and ATLAS images extended the orbit to mid-June, confirming extreme eccentricity, $e \approx 6$ \citep{Seligman25}. Early follow-up showed 3I/ATLAS was already active at $r \simeq 4.4$~au, with a dust coma detected by multiple facilities \citep{Bolin25,deLaFuenteMarcos25}; TESS \citep{Feinstein2025,MartinezPalomera2025} and Rubin/LSST \citep{Chandler2025} precovery data further extend the observational arc. 

JPL Horizons \citep{Horizons25} predicts perihelion on 29–30 October 2025 at $q \approx 1.36$~au, just inside Mars’ orbit, before it recedes into interstellar space. Ground-based observations will be impossible from September due to low solar elongation, with the object reappearing in the northern sky in late November and remaining visible until mid-2026.

Here, we present optical photometry and low-resolution spectroscopy of 3I/ATLAS obtained during its first month of observations, still beyond 3.5~au, establishing a pre-perihelion baseline for tracking its evolution.

%%%%%%%%%%%%%%%%%%%%%%%%%%%%%%%%%%%%%%%%%%%%%%%%%%%%%%%%%%%%%%
\section{Observations}

\subsection{Spectroscopy}

\begin{figure}
\centering
\includegraphics[angle=0,width=0.5\textwidth]{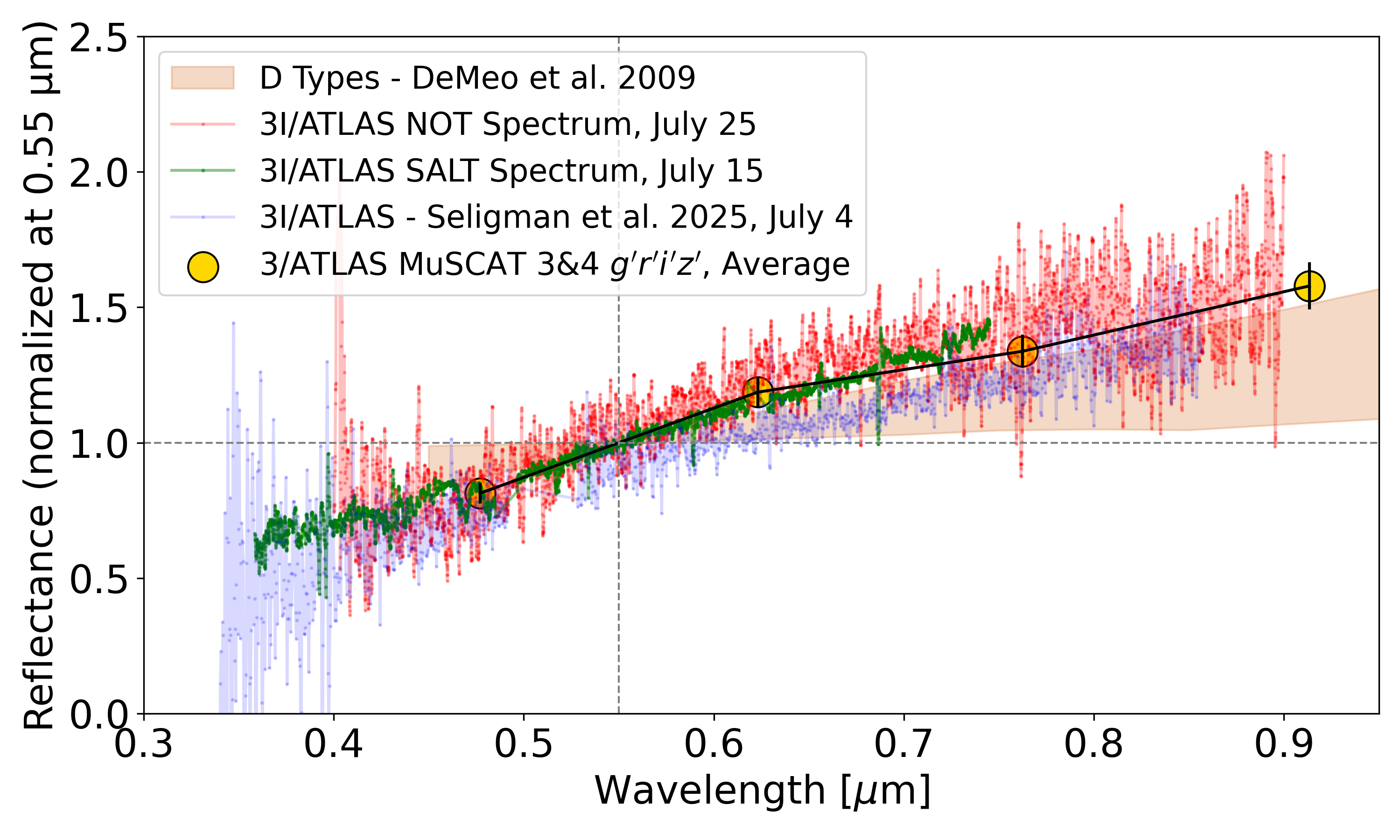}\\
\caption{Comparison of the spectrum of 3I/ATLAS obtained with the Southern African Large Telescope (SALT) and the Nordic Optical Telescope (NOT) against the reference spectrum reported by \citet{Seligman25}. Overlaid on the spectral data are the campaign-averaged $g'r'i'z'$ photometric fluxes observed with the Faulkes Telescope North (FTN). The D-type asteroid spectrum \citep{DeMeo2009} is plotted for reference. All the datasets are normalized to 0.55 $\mu\mathrm{m}$.}
\label{fig:spectra}
\end{figure}

Low-resolution spectroscopic observations of 3I/ATLAS (hereafter 3I) were obtained on 15 and 25 July using the 10-meter Southern African Large Telescope (SALT; see Section~\ref{sec:appendix_salt}) and the 2.56-meter Nordic Optical Telescope (NOT; see Section~\ref{sec:appendix_not}), respectively. The SALT spectrum spanned $0.36$--$0.74~\mu\mathrm{m}$, while the NOT spectrum covered $0.4$--$0.90~\mu\mathrm{m}$. To remove the solar continuum and derive the comet’s reflectance spectrum, we observed the solar analogue SA112-1333 \citep{landolt} on the same nights. Spectra of the spectrophotometric standard Feige~110 \citep{oke1990} were also obtained for flux calibration. Standard data reduction procedures were applied, after which individual comet spectra were median-combined and divided by the solar analogue. The resulting reflectance spectra are shown in Fig.~\ref{fig:spectra}.

\subsection{Photometry}

Photometric observations of 3I were carried out between 2 and 29 July 2025 using the Faulkes Telescopes North and South (FTN and FTS; see Section~\ref{sec:appendix_ftn}), the Nordic Optical Telescope (NOT; see Section~\ref{sec:appendix_not_ph}), the Telescopi Joan Oró (TJO; see Section~\ref{sec:appendix_tjo}), the Lesedi Telescope (see Section~\ref{sec:appendix_lesedi}), and the Skalnaté Pleso Observatory (see Section~\ref{sec:appendix_spo}). Time-series photometry was performed in the SDSS $r'$ band, while multi-band imaging in $g'r'i'z'$ and VR filters provided color measurements. Exposure times were adjusted dynamically according to seeing and plate scale, maximizing integration while keeping stellar point sources untrailed. All images were calibrated with standard procedures, including bias subtraction, dark-frame correction, and flat-fielding. Photometry was obtained via aperture photometry relative to selected field stars, with calibration based on reference magnitudes from the ATLAS All-Sky Stellar Reference Catalog \citep{TonryJ2018} or the APASS catalog \citep{henden2014apass}.

\section{Results}

\subsection{Analysis of the color evolution}

\begin{figure}
\centering
\includegraphics[angle=0,width=0.5\textwidth]{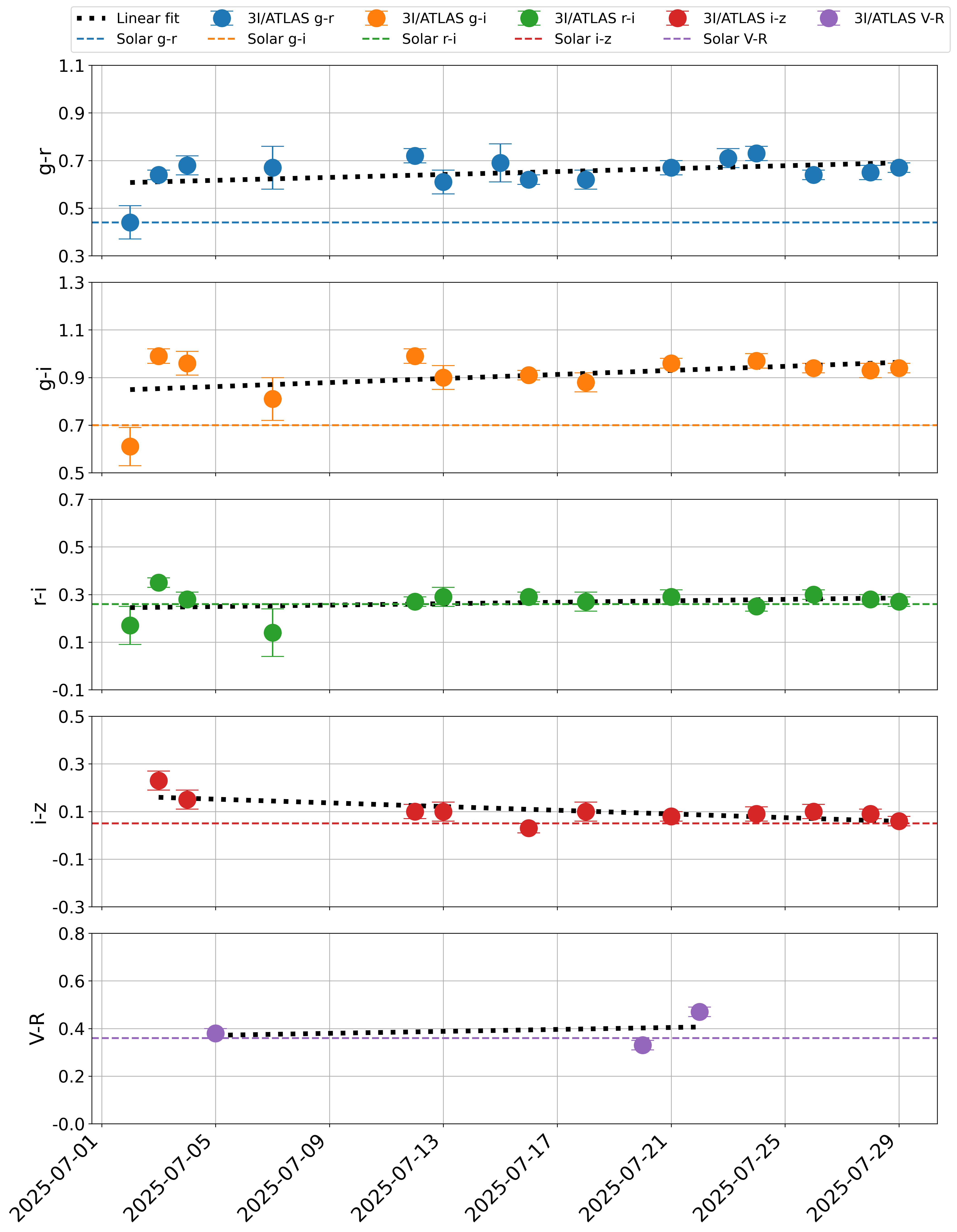}
\caption{Evolution of photometric colors over the campaign from 02-29 July. A reddening trend is observed in the wavelength range covered by the g, r, and i bands, while a bluing trend is seen in the i–z color.}
\label{fig:colors}
\end{figure}

In the $0.4$–$0.7~\mu\mathrm{m}$ range we observed slight reddening as the comet approached the Sun (rising $g-r$ and $g-i$ in Fig.~\ref{fig:colors}), a neutral trend at $0.7$–$0.8~\mu\mathrm{m}$ (constant $r-i$), and slight bluing beyond $0.8~\mu\mathrm{m}$ (decreasing $i-z$). Overall, the color is bluer than reported by \citet{Bolin25} for 3I at similar distances, but closer to \citet{belyakov2025spectral} and \citet{opitom2025initial}, who also found neutral to slightly red values. Compared to other bodies, 3I’s color resembles that of 2I/Borisov near 2~au \citep{epifani2021colour} and distant comets near 4~au (e.g., \citealt{snodgrass2008optical}; \citealt{ivanova2015observations}; \citealt{voitko2024dust}), suggesting analogous dust properties.

Spectroscopy shows a similar trend: spectral gradients from NOT and SALT reflectance data indicate progressive reddening. \citet{Seligman25} reported $17.1 \pm 0.2~\%/\mu\mathrm{m}$ on 4~July, while we measure $21.1 \pm 0.2~\%/\mu\mathrm{m}$ (SALT, 15~July) and $22.8 \pm 0.1~\%/\mu\mathrm{m}$ (NOT, 25~July) over $0.4$–$0.7~\mu\mathrm{m}$. Beyond $0.7~\mu\mathrm{m}$, the NOT spectrum flattens to $9.8 \pm 0.3~\%/\mu\mathrm{m}$ at $0.7$–$0.9~\mu\mathrm{m}$, consistent with \citealt{opitom2025initial} and \citealt{deLaFuenteMarcos25}. In contrast, NIR spectra \citep{Kareta2025,Yang2025} show strong flattening beyond $0.9~\mu$m, with slopes $\sim3~\%/1000~\AA$, implying a transition to neutral or slightly blue slopes. Thus, 3I shows wavelength-dependent evolution: redder in the optical, flatter or bluer in the NIR, consistent with the $i-z$ color trend. Future spectroscopy will track slope changes, emission lines, and activity.

\subsection{Magnitude and dust production level}

Between 2 and 29 July, photometric observations of 3I show a progressive brightening, consistent with its decreasing heliocentric distance from 4.5~au to $\sim3.6$~au. The absolute magnitude was $H \sim 12$~mag, in agreement with the $H \sim 11.9$ reported by \citet{jewitt2025interstellar}. Assuming a geometric albedo $p_v = 0.04$, the estimated upper limit on the comet radius is $\sim 11$~km (see Table~\ref{tab:merged_table}). The dust activity level, expressed as $Af\rho$, is $\sim 300$~cm, consistent with \citet{Bolin25}, but higher than values observed for 2I/Borisov (\citealt{guzik2020initial}; \citealt{opitom20192i}). This is comparable to measurements of Jupiter-family comets at large heliocentric distances and some distant comets exhibiting similar activity (\citealt{ivanova2015observations}; \citealt{voitko2024dust}). Estimated dust mass-loss rates range from $\sim 0.3$ to $4.2$~kg~s$^{-1}$, slightly above those reported by \citet{Bolin25}, yet lower than typical rates for Jupiter-family comets at large distances \citep{gillan2024dust}, and comparable to distant comets observed even farther from the Sun \citep{ivanova2015observations}. Detailed methodology, relevant equations, and complete tables are provided in Appendix~\ref{sec:appendix_photometry}.

\subsection{Lightcurve}

\begin{figure}
\centering
\includegraphics[angle=0,width=0.99\columnwidth]{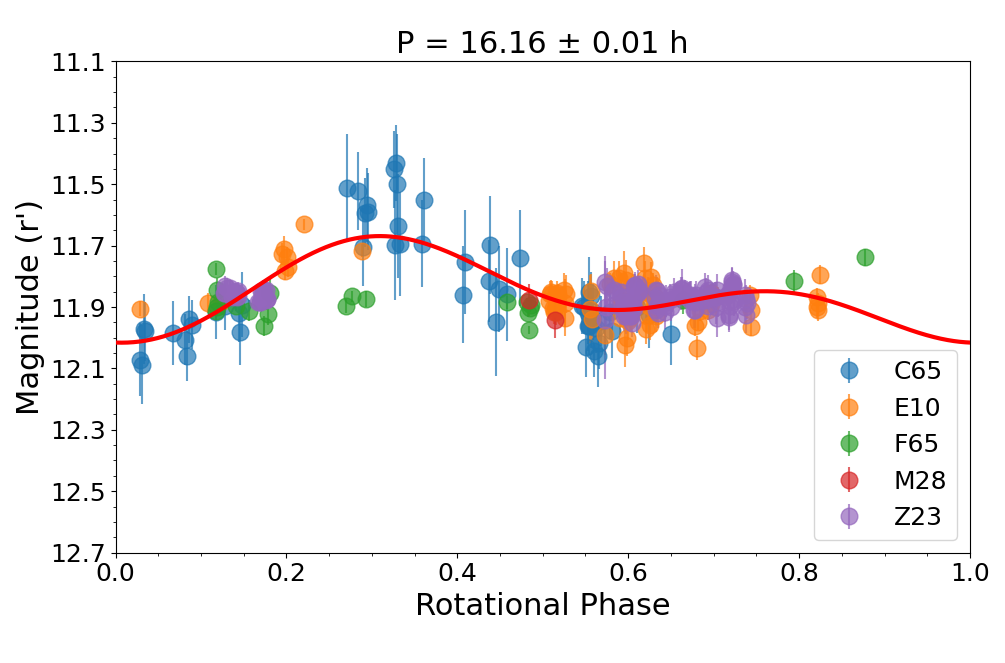}   
\caption{Rotational lightcurve of comet 3I/ATLAS phased with a spin period of $16.16 \pm 0.01$~h. Photometric data points from multiple telescopes are shown with error bars; their Minor Planet Center (MPC) codes are listed in the box within the figure. The red solid line represents the best-fitting second-order Fourier model to the phased lightcurve. The MPC codes correspond to the following telescopes: C65 (Telescopi Joan Oró), E10 (Faulkes Telescope South), F65 (Faulkes Telescope North), M28 (Lesedi Telescope), and Z23 (Nordic Optical Telescope).}
\label{fig:lc}
\end{figure}

Using the time-series photometry described in Section~2.2, we generated a rotational lightcurve of 3I. Observations occurred while the comet crossed the Galactic plane, requiring mitigation techniques to minimize stellar contamination. Fourier analysis yielded a best-fitting spin period of $16.16 \pm 0.01$~h (see Figure~\ref{fig:lc}), with a peak-to-peak amplitude of $\sim0.3$~mag and prominent aliases near 8 and 24~h (see Figure~\ref{fig:periodogram}). This broadly agrees with the $\sim16.79$~h period reported by \citet{deLaFuenteMarcos25}, though their estimate—based on 2--5 July photometry—may be affected by the limited temporal baseline.

The lightcurve amplitude $A$ can be related to the projected axis ratio of a triaxial ellipsoid via $b/a = 10^{-0.4A}$ \citep[e.g.,][]{Binzel1989}, giving $b/a \approx 0.76$–0.83 for $A = 0.2$–0.3~mag, which represents minimum elongations for equatorial viewing and small phase angles ($\alpha = 4^\circ$–$12^\circ$).

Using the Hubble Space Telescope (HST) upper limit on the nucleus radius, \(r_n \le 2.8\)~km \citep{Jewitt2025}, the maximum principal axes are
\begin{equation}
a_\mathrm{max} \approx \frac{r_n}{\sqrt{b/a}} \sim 3.1\text{--}3.2~\mathrm{km}, \quad
b_\mathrm{max} \approx r_n \sqrt{b/a} \sim 2.4\text{--}2.6~\mathrm{km}.
\end{equation}

\section{Conclusions}

The morphology of comet 3I over one month reveals clear activity, with continuous dust emission evidenced by the increasing coma size as the heliocentric distance decreased. The absence of a visible tail likely reflects the observing geometry. The coma remained predominantly asymmetric, with no signs of fragmentation or rapid changes. Overall, 3I shows characteristics typical of weakly active outer Solar System comets, despite its interstellar origin. Time-series photometry indicates a spin period of $16.16 \pm 0.01$~h, with a maximum lightcurve amplitude of 0.3~mag at the start, decreasing to 0.2~mag—likely as increasing activity progressively masks the nucleus's rotational signal. The lightcurve amplitude $A$ relates to the projected axis ratio via $b/a = 10^{-0.4A}$ \citep[e.g.,][]{Binzel1989}, giving $b/a \approx 0.76$--0.83 for $A = 0.2$--0.3~mag, representing minimum elongations for equatorial viewing and small phase angles ($\alpha = 4^\circ$--$12^\circ$). Enhanced activity is further supported by a reddening trend in the comet’s colors.

Measured $Af\rho$ values ($\sim$300~cm) are comparable to distant Jupiter-family comets \citep{gillan2024dust}, but dust production rates are lower than expected for an object of this brightness at $r \simeq 4$~au, suggesting a coma dominated by large particles. Independent HST observations report dust mass-loss rates of $\sim$6--60~kg~s$^{-1}$ \citep{Jewitt2025}, also indicating large grains. The absence of a distinct tail and the homogeneous coma morphology support this, as massive particles are less affected by solar radiation pressure and remain near the nucleus, implying a deficit of submicron dust that would otherwise enhance light scattering. Further polarimetric and infrared observations are needed to constrain the particle size distribution. We plan to continue photometric and spectroscopic monitoring, including polarimetry to detect changes in color and light-scattering properties of the coma pre- and post-perihelion. These observations are essential to characterize interstellar material, its evolution under solar radiation, and to enable comparisons with Solar System small bodies.

%%%%%%%%%%%%%%%%%%%%%%%%%%%%%%%%%%%%%%%%%%%%%%%%%%%%%%%%%%%%%%
\begin{acknowledgements}

T.S-R.\ acknowledges funding from Ministerio de Ciencia e Innovaci{\'o}n (Spanish Government), PGC2021, PID2021-125883NB-C21. This work was (partially) supported by the Spanish MICIN/AEI/10.13039/501100011033 and by ``ERDF A way of making Europe" by the “European Union” through grant PID2021-122842OB-C21, and the Institute of Cosmos Sciences University of Barcelona (ICCUB, Unidad de Excelencia `Mar\'ia de Maeztu’) through grant CEX2019-000918-M. D.O.\ was supported by grant No.\ 2022/45/B/ST9/00267 from the National Science Centre, Poland. O.I.\ and M.H.\ thank the Slovak Grants by the Agency for Science, VEGA, grant No.\ 2/0059/22.

This work is based on observations made at the South African Astronomical Observatory (SAAO), which is financially supported by the South African National Research Foundation (NRF). Some of the observations reported in this paper were obtained with the Southern African Large Telescope (SALT). Polish participation in SALT is funded by grant No.\ MEiN nr 2021/WK/01. The Joan Or\'{o} Telescope (TJO) at the Montsec Observatory (OdM) is owned by the Catalan Government and operated by the Institute of Space Studies of Catalonia (IEEC). Partly based on observations made with the Nordic Optical Telescope. The NOT data were obtained under program ID 71-411 (PI A. Penttil\"a).

Work of Skalnat\'{e} Pleso has made use of data from the Asteroid Terrestrial-impact Last Alert System (ATLAS) project. ATLAS is primarily funded to search for near-Earth asteroids through NASA grants NN12AR55G, 80NSSC18K0284, and 80NSSC18K1575; byproducts of the NEO search include images and catalogs from the survey area. The ATLAS science products have been made possible through the contributions of the University of Hawaii Institute for Astronomy, the Queen's University Belfast, the Space Telescope Science Institute, and the South African Astronomical Observatory.

This paper is based on observations made with the MuSCAT instruments, developed by the Astrobiology Center (ABC) in Japan, the University of Tokyo, and Las Cumbres Observatory (LCOGT). MuSCAT3 was developed with financial support by JSPS KAKENHI (JP18H05439) and JST PRESTO (JPMJPR1775), and is located at the Faulkes Telescope North on Maui, HI (USA), operated by LCOGT. MuSCAT4 was developed with financial support provided by the Heising-Simons Foundation (grant 2022-3611), JST grant number JPMJCR1761, and the ABC in Japan, and is located at the Faulkes Telescope South at Siding Spring Observatory (Australia), operated by LCOGT.

Some observations were obtained by the Comet Chasers schools outreach program (https://www.cometchasers.org/), which is funded by the UK Science and Technology Facilities Council (via the DeepSpace2DeepImpact Project), the Open University, and Cardiff University. It accesses the LCOGT telescopes through the Schools Observatory/Faulkes Telescope Project (TSO2025A-00 DFET—The Schools' Observatory), which is partly funded by the Dill Faulkes Educational Trust, and the LCO Global Sky Partners Programme (LCOEPO2023B-013). Schools making observations included The Coopers Company \& Coborn School, Upminster, UK, Ysgol Gyfun Gymraeg Bro Edern, Cardiff, UK, St Marys Catholic Primary School, Bridgend, UK, Institut d'Alcarràs, Catalonia, Spain, Louis Cruis Astronomy Club and Srednja škola Jelkovec (Jelkovec High School), Zagreb, Croatia.

\end{acknowledgements}

\bibliography{3I-ATLAS.bib} % your references Yourfile.bib

% %%%%%%%%%%%%%%%%%%%%%%%%%%%%%%%%%%%%%%%%%%%%%%%%%%%%%%%%%%%%%%
% Example below of non-structurated natbib references  
% To use the v8.3 macros with this form of composition of bibliography,
% the option "bibyear" should be added to the command line
% "\documentclass[bibyear]{aa}".
% %%%%%%%%%%%%%%%%%%%%%%%%%%%%%%%%%%%%%%%%%%%%%%%%%%%%%%%%%%%%%%

% \begin{thebibliography}{}

%   \bibitem[1966]{baker} Baker, N. 1966,
%       in Stellar Evolution,
%       ed.\ R. F. Stein,\& A. G. W. Cameron
%       (Plenum, New York) 333

%    \bibitem[1988]{balluch} Balluch, M. 1988,
%       A\&A, 200, 58

%    \bibitem[1980]{cox} Cox, J. P. 1980,
%       Theory of Stellar Pulsation
%       (Princeton University Press, Princeton) 165

%    \bibitem[1969]{cox69} Cox, A. N.,\& Stewart, J. N. 1969,
%       Academia Nauk, Scientific Information 15, 1

%    \bibitem[1980]{mizuno} Mizuno H. 1980,
%       Prog. Theor. Phys., 64, 544
   
%    \bibitem[1987]{tscharnuter} Tscharnuter W. M. 1987,
%       A\&A, 188, 55
  
%    \bibitem[1992]{terlevich} Terlevich, R. 1992, in ASP Conf. Ser. 31,
%       Relationships between Active Galactic Nuclei and Starburst Galaxies,
%       ed. A. V. Filippenko, 13

%    \bibitem[1980a]{yorke80a} Yorke, H. W. 1980a,
%       A\&A, 86, 286

%    \bibitem[1997]{zheng} Zheng, W., Davidsen, A. F., Tytler, D. \& Kriss, G. A.
%       1997, preprint
% \end{thebibliography}

%%%%%%%%%%%%%%%%%%%%%%%%%%%%%%%%%%%%%%%%%%%%%%%%%%%%%%%%%%%%%%%
% Appendices must be placed after   \end{thebibliography}
% They will be placed automatically on a new page.
%%%%%%%%%%%%%%%%%%%%%%%%%%%%%%%%%%%%%%%%%%%%%%%%%%%%%%%%%%%%%%%
\begin{appendix}
%%%%%%%%%%%%%%%%%%%%%%%%%%%%%%%%%%%%%%%%%%%%%%%%%%%%%%%%%%%%%%%
% In the PDF output, floats should be placed
% under their own appendix, not before the title, nor after the
% title of the next appendix.

% In short appendices, onecolumn floats (\figure*
% or \table*) will generate a blank page.
% To prevent this behaviour, a few examples are provided here. 

% In case you have a lot of floating objects for little text and the 
% LaTeX engine moves the floats away from their context, the command
% \FloatBarrier of the “placeins” package will empty the
% float buffer and place all stored floats in the continuity.

% If you still encounter problems with wide floats placement,
% just use the onecolumn environment throughout the appendices.
%%%%%%%%%%%%%%%%%%%%%%%%%%%%%%%%%%%%%%%%%%%%%%%%%%%%%%%%%%%%%%%

%____________________________________________________________
%       Wide floats at the start of an appendix: first method
%-------------------------------------------------------------
% To prevent a blank page after the start of an appendix:
% - Switch to one \onecolumn first
% - Declare the section title
% - Declare the onecolumn float with the parameter [h!]
% - Revert to \twocolumn at the end of the section

\section{Observational facilities}

\subsection{Photometry}

\subsubsection{Faulkes Telescopes North and South}
\label{sec:appendix_ftn}

We used the 2.0-m Faulkes Telescope North (FTN; MPC code F65), located at Haleakalā Observatory in Maui, Hawai‘i (USA), and the 2.0-m Faulkes Telescope South (FTS; MPC code E10), located at Siding Spring Observatory in New South Wales (Australia), to obtain photometry in the SDSS $g'$, $r'$, $i'$, and $z'$ bands using the MuSCAT3 and MuSCAT4 instruments, respectively. MuSCAT3 is equipped with four 1k~$\times$~1k CCDs (Hamamatsu Photonics), each with a pixel scale of 0.27\arcsec\ and a field of view of 7.4~$\times$~7.4~arcmin$^2$, while MuSCAT4 features four 2k~$\times$~2k CCDs, with a pixel scale of 0.38\arcsec\ and a field of view of 13~$\times$~13~arcmin$^2$ \citep{Narita2020, Fukui2022}. Both instruments enable simultaneous multi-band imaging via dichroic beam splitters. The detectors exhibit low readout noise ($\sim$3--5~e$^-$), and the photometric data were processed using the MuSCAT pipeline, which includes standard calibrations (bias, dark, and flat-field corrections) and aperture photometry optimized for each band.

\subsubsection{Nordic Optical Telescope}
\label{sec:appendix_not_ph}

Photometric observations in the Sloan $r'$ filter were performed at the Nordic Optical Telescope (NOT) using the Alhambra Faint Object Spectrograph and Camera (ALFOSC). The NOT is a 2.56-meter telescope located at the Spanish ``Roque de los Muchachos'' Observatory (ORM), La Palma, Canarias, Spain (MPC code: 950). On 17 July 2025, three series of 80 exposures with 30-second integration times were obtained, and on 18 July 2025, one series of 80 exposures with 30-second integration times was acquired.

\subsubsection{Telescopi Joan Or\'{o}}
\label{sec:appendix_tjo}

We used the 0.8-m Telescopi Joan Oró (TJO; MPC code C65), located in the Catalan Pyrenees, to obtain time-series photometry in the SDSS $r'$ band on the nights of 4, 6, 13, 16, 17, and 19 July. On 4 and 6 July, the backup camera MEIA2 (2k~$\times$~2k Andor iKon XL, pixel scale 0.36\arcsec) was used, while on the remaining nights the main camera MEIA3 (4k~$\times$~4k Andor iKon XL 230-84, pixel scale 0.4\arcsec) was employed. Both cameras have very similar performance, differing slightly in gain (1.55~e$^-$/ADU for MEIA2 vs.\ 1.04~e$^-$/ADU for MEIA3) and readout noise (8~e$^-$ vs.\ 9~e$^-$, respectively). The main distinction lies in the field of view: 0.13~deg for MEIA2 and 0.45~deg for MEIA3. All images were calibrated using the IEEC Calibration and Analysis Tool (ICAT; \citealt{colome2010}).

\subsubsection{Lesedi}
\label{sec:appendix_lesedi}
Multi-filter photometric observations were collected with the f/8, 1-meter Lesedi telescope \citep{Worters2016}, located at the South African Astronomical Observatory’s (SAAO) observing site in the Northern Cape, South Africa (GPS coordinates: 32°22$^{\prime}$47.2$^{\prime\prime}$\,S, 20°48$^{\prime}$38.4$^{\prime\prime}$\,E; MPC code: M28). The instrument used was Mookodi \citep{Erasmus2024a}, which in imaging mode provides a field of view of 10\,arcmin\,$\times$\,10\,arcmin and is equipped with a full SDSS filter set ($u'g'r'i'z'$). The telescope operates in a fully robotic mode each night as part of the SAAO’s ``Intelligent Observatory'' (IO) initiative \citep{Potter2024, Erasmus2024b, Erasmus2025}, and data were collected under the rapid-response near-Earth asteroid follow-up program currently running at the SAAO (PIs: T.S Ngwane and N. Erasmus).

Observations were obtained on 02, 07, 15 and 23 July 2025 using 1$\times$1 binning (0.59 arcsec/pixel), the slowest readout mode of 0.5\,MHz (to minimize readout noise), and the highest gain setting of 1.2\,e$^-$/ADU.

\begin{figure}
\centering
\includegraphics[angle=0,width=0.8\columnwidth]{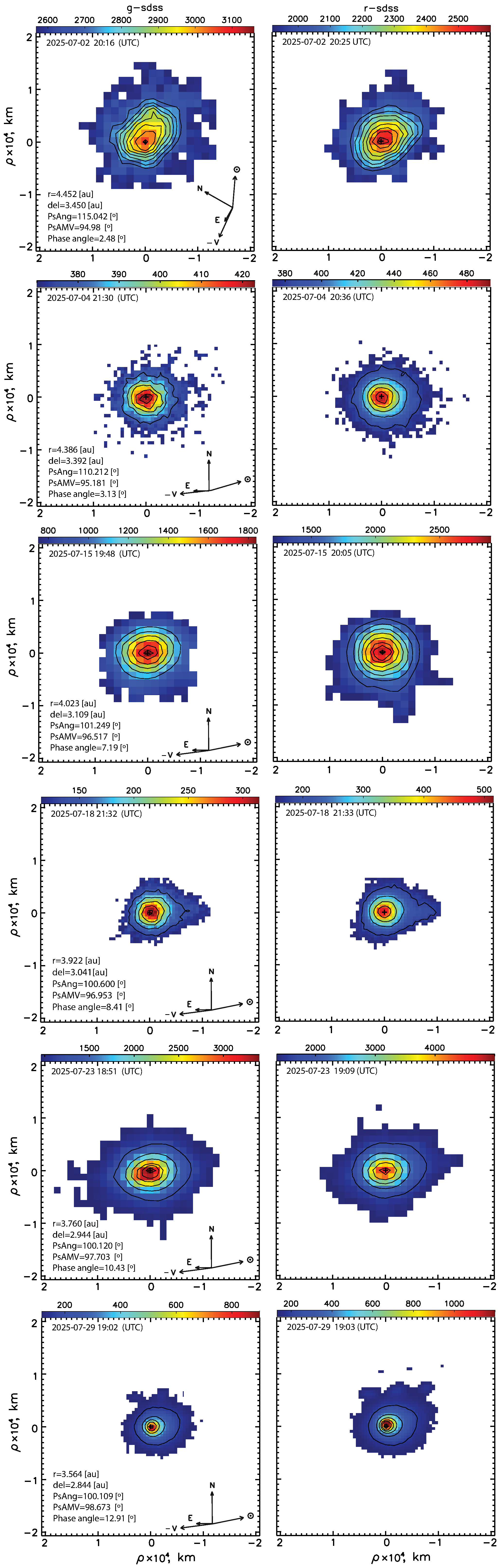}   
\caption{Intensity maps of comet 3I/ATLAS in the $g$ and $r$ SDSS filters. The color scale does not reflect the absolute brightness of the comet. Arrows indicate the directions toward the Sun, celestial north, east, and the negative velocity vector of the comet projected onto the plane of the sky.
\label{fig:morphol}}
\end{figure}

On 02/07, a time-series observation was carried out by cycling through the $g'$, $r'$, $i'$, and $z'$ filters continuously between 20:06:02 and 20:45:05, with an exposure time of 120 seconds per filter. The five $g'$-filter and five $r'$-filter frames collected during this window were median-stacked and used to generate the intensity maps shown in the top row of Fig.\,\ref{fig:morphol}.

On 07/07, two time-series datasets were obtained. The first was in the $z'$ filter only, spanning approximately one hour between 19:35:27 and 20:20:59, using 30-second exposures. The second followed a similar multi-filter sequence to that of 02/07, with 30-second exposures between 20:26:18 and 21:15:12. The seven $g'$-filter and six $r'$-filter frames collected during this window were median-stacked.

On 15/07, a time-series cycling between only the $g'$ and $r'$ filters was performed, with 30-second exposures between 19:48:07 and 20:42:47. Approximately 20 frames per filter were collected and stacked to generate the intensity maps shown in the third row of Fig.\,\ref{fig:morphol}. 

A near-identical observing strategy to that used on 15/07 was employed on 23/07, and the resulting analysis is shown in the fifth row of Fig.\,\ref{fig:morphol}. 

The photometric colors, dust activity levels, and dust mass-loss rates were also extracted from the median stacks in the relevant filters for each days and are reported in Tables\,\ref{tab:table_phot_combined} and \ref{tab:merged_table}.

\subsubsection{Skalnat\'{e} Pleso Observatory}
\label{sec:appendix_spo}

Photometric material at the Skalnat\'{e} Pleso Observatory (MPC code: 056) for the interstellar comet 3I/ATLAS was acquired using a 0.61-m f/4.3 Newtonian telescope equipped with an SBIG ST-10XME CCD camera and Johnson-Cousins $R$ and $V$ photometric filters. Observations were conducted using $2 \times 2$ binning, yielding a resolution of 1.07~arcsec per pixel. Standard reduction procedures were applied, including bias subtraction, dark frame subtraction, and flat-field correction. Tycho Tracker software \citep{ParrottD2020} was used as the main tool for deriving magnitudes and $Af\rho$ quantities. Stellar magnitudes for standard stars were taken from the ATLAS catalog, an all-sky reference catalog containing approximately one billion stars down to a limiting magnitude of 19. The catalog integrates data from Pan-STARRS DR1, ATLAS Pathfinder, ATLAS re-flattened APASS, SkyMapper DR1, APASS DR9, Tycho-2, and the Yale Bright Star Catalog. Developed by \citet{TonryJ2018}, it provides a robust foundation for photometric calibration \citep{KostovA2018}. Interstellar comet 3I/ATLAS was observed on five nights between July 3 and July 23.

\subsection{Spectroscopy}

\subsubsection{Southern African Large Telescope}
\label{sec:appendix_salt}

Low-resolution spectroscopy of 3I/ATLAS was performed using the 10-meter Southern African Large Telescope (SALT; \citealt{Kwiatkowski2009}), located in Sutherland, South Africa (IAU code: B31). Observations were conducted on 2025 July 15 using the Robert Stobie Spectrograph (RSS; \citealt{Burgh2003, Kobulnicky2003}) under program 2025-1-DDT-004 (PI: T. Kwiatkowski). The RSS is equipped with a detector comprising a $3 \times 1$ mini-mosaic of CCD chips, each with $2048 \times 4102$ pixels of 15~$\mu$m size and a pixel scale of 0.12~arcseconds per unbinned pixel. We obtained 20 exposures of 120~s each using the PG0700 grating and a $3''$ narrow slit, aligned along the comet's trail. Observations were conducted in $4 \times 4$ binning mode.

Due to the comet’s proximity to the Milky Way, significant contamination from background stars occurred along the slit. Consequently, only 5$\times$120~s exposures, during which the comet was positioned against a clean sky region, were retained. All spectra obtained along the slit were extracted, but only the comet spectrum was analyzed. A solar analogue star, SA112-1333, was observed on the same night and at a similar airmass for calibration purposes.

Data reduction and calibration followed standard procedures, including bias subtraction, flat-field correction, sky background subtraction, and one-dimensional aperture extraction. Wavelength calibration was performed using an argon lamp spectrum. The individual spectra of 3I/ATLAS were median-combined and divided by the median-combined spectrum of the solar analogue to remove the solar continuum. The final reflectance spectrum spans the wavelength range 3600–7400~\AA, with gaps between 4820–4980~\AA\ and 6160–6300~\AA.

\subsubsection{Nordic Optical Telescope}
\label{sec:appendix_not}

Long-slit spectroscopic observations were carried out at the Nordic Optical Telescope (NOT) using ALFOSC on 25 July 2025, at an airmass of approximately $\sim$1.5. A series of 13~$\times$~180-second exposures were obtained using Grism~\#4, which covers the wavelength range from 0.32~\textmu m to 0.96~\textmu m, with a 1.8-arcsecond-wide slit. To minimize atmospheric dispersion, the instrument was aligned along the parallactic angle. No second-order blocking filter was used. 

It should be noted that for Grism~\#4, second-order light from the $U$-band becomes noticeable above 5900~\AA, contributing up to 20\% of the recorded signal, and from the $B$-band contributes around 5\%. For flux calibration and spectral normalization, the solar analogue BD-004074 and the spectrophotometric standard star Feige~110 were observed using the same setup.

We followed standard data reduction procedures. Helium and neon arc lamps were used for wavelength calibration, while halogen flats were employed for pixel response correction. Bias frames were also acquired to correct for the detector’s electronic offset.

\begin{table*}
    \centering
    \caption{The results of the multi-band photometry of 3I/ATLAS by Faulkes Telescope North (FTN), Faulkes Telescope South (FTS), Lesedi, and Skalnat\'{e} Pleso telescopes.}
    \label{tab:table_phot_combined}
    \begin{tabular}{c c c c c c c c c}
    \hline
    \makecell{Date\\(2025)} & Telescope & $\rho^{*}$, km & $m_\mathrm{r}$, mag & $g-r$, mag & $g-i$, mag & $r-i$, mag & $i-z$, mag & $V-R$, mag \\
    \hline
    Jul 2  & Lesedi         & 10352 & 17.85$\pm$0.05 & 0.44$\pm$0.07 & 0.61$\pm$0.08 & 0.17$\pm$0.08 & --            & -- \\
    Jul 3  & FTN        & --    & 18.08$\pm$0.02 & 0.64$\pm$0.02 & 0.99$\pm$0.03 & 0.35$\pm$0.02 & 0.23$\pm$0.04 & -- \\
%    Jul 3  & Skalnaté Pleso & 10606 & --             & --            & --            & --            & --            & -- \\
    Jul 4  & FTS        & --    & 18.08$\pm$0.02 & 0.68$\pm$0.04 & 0.96$\pm$0.05 & 0.28$\pm$0.03 & 0.15$\pm$0.04 & -- \\
    Jul 5  & Skalnaté Pleso & 10418 & --             & --            & --            & --            & --            & 0.38$\pm$0.02 \\
    Jul 7  & Lesedi         & 10634 & 17.31$\pm$0.07 & 0.67$\pm$0.09 & 0.81$\pm$0.09 & 0.14$\pm$0.10 & --            & -- \\
    Jul 12 & FTN        & --    & 17.94$\pm$0.01 & 0.72$\pm$0.03 & 0.99$\pm$0.03 & 0.27$\pm$0.02 & 0.10$\pm$0.03 & -- \\
    Jul 13 & FTN        & --    & 17.86$\pm$0.02 & 0.61$\pm$0.05 & 0.90$\pm$0.05 & 0.29$\pm$0.04 & 0.10$\pm$0.04 & -- \\
    Jul 15 & Lesedi         & 10661 & 17.20$\pm$0.05 & 0.69$\pm$0.08 & --            & --            & --            & -- \\
    Jul 16 & FTN        & --    & 17.96$\pm$0.01 & 0.62$\pm$0.02 & 0.91$\pm$0.02 & 0.29$\pm$0.02 & 0.03$\pm$0.02 & -- \\
    Jul 18 & FTN/FTS        & --    & 17.80$\pm$0.02 & 0.62$\pm$0.04 & 0.88$\pm$0.04 & 0.27$\pm$0.04 & 0.10$\pm$0.04 & -- \\
    Jul 20 & Skalnaté Pleso & 11640 & --             & --            & --            & --            & --            & 0.33$\pm$0.02 \\
    Jul 21 & FTN        & --    & 17.78$\pm$0.02 & 0.67$\pm$0.03 & 0.96$\pm$0.02 & 0.29$\pm$0.03 & 0.08$\pm$0.02 & -- \\
    Jul 22 & Skalnaté Pleso & 11490 & --             & --            & --            & --            & --            & 0.47$\pm$0.02 \\
    Jul 23 & Lesedi         & 10584 & 17.15$\pm$0.02 & 0.71$\pm$0.04 & --            & --            & --            & -- \\
    Jul 23 & Skalnaté Pleso         & 11418 & -- & --            & --            & --            & -- & 0.51$\pm$0.02 \\
    Jul 24 & FTS        & --    & 17.52$\pm$0.01 & 0.73$\pm$0.03 & 0.97$\pm$0.03 & 0.25$\pm$0.02 & 0.09$\pm$0.03 & -- \\
    Jul 26 & FTN        & --    & 17.71$\pm$0.01 & 0.64$\pm$0.02 & 0.94$\pm$0.02 & 0.30$\pm$0.02 & 0.10$\pm$0.03 & -- \\
    Jul 28 & FTN        & --    & 17.55$\pm$0.01 & 0.65$\pm$0.03 & 0.93$\pm$0.03 & 0.28$\pm$0.02 & 0.09$\pm$0.02 & -- \\
    Jul 29 & FTN        & --    & 17.73$\pm$0.01 & 0.67$\pm$0.02 & 0.94$\pm$0.02 & 0.27$\pm$0.02 & 0.06$\pm$0.02 & -- \\

    \hline

    \textbf{Average} & -- & -- & -- 
    & \textbf{0.65$\pm$0.03} & \textbf{0.91$\pm$0.03} & \textbf{0.27$\pm$0.03} & \textbf{0.10$\pm$0.04} & \textbf{0.42$\pm$0.02} \\
    \hline
\multicolumn{9}{l}{\footnotesize $^*$ The projected size of the photometric aperture at the distance of the comet.} \\
    \end{tabular}
\end{table*}

\section{Morphology of the coma} 

We created intensity maps (Fig.~\ref{fig:morphol}) by co-adding the photometric images taken with the SDSS $g'$ and $r'$ filters on the two dates of the observing campaign, 4 and 29 July 2025. A more detailed view of the comet’s morphological evolution from 2 to 29 July is presented in Fig.~\ref{fig:morphol}. The cometary coma appears compact and slightly asymmetrical. No visible tail was detected before 16 July. A slight elongation of the coma, resembling a short tail, was observed between 18 and 26 July, directed roughly at a position angle of $280^{\circ}$. Observations by \citet{kareta2025near} using the NASA Infrared Telescope Facility similarly showed no prominent tail. This absence, also evident in our images and those from other observers, can be partially attributed to the tail being projected ‘behind’ the comet due to the low phase angle during the observations. This effect may be further enhanced if the comet contains relatively few fine dust grains, which are the most affected by radiation pressure. To search for faint structures within the weak coma of comet 3I, we applied digital image enhancement techniques, including the rotational gradient method \citep{larson1984coma}, the 1/$\rho$ profile, and renormalization filtering \citep{samarasinha2014image}. These methods were applied only to the stacked images, as the object was too faint for filtering of individual frames. No jet-like or other morphological features were detected.

The absence of distinct morphological structures in comet 3I is consistent with its currently low level of activity. Long-period and dynamically new comets can display enhanced activity \citep{Meech2004}, with noticeable coma structures even at large heliocentric distances \citep{Jewitt19,Ivanova2019,Ivanova2021,Ivanova2023}. However, such behavior strongly depends on the intrinsic properties of the nucleus, including its volatile inventory and the spatial distribution of active areas. For comparison, comet 2I/Borisov exhibited much stronger activity, characterized by an elongated coma, jet-like outflows, and a well-developed tail \citep{Manzini2020}. The limited morphology observed in comet 3I therefore reflects both its early activation stage and its particular physical characteristics.

\section{Photometry and dust parameters derivation}
\label{sec:appendix_photometry}

\begin{table*}
    \centering
    \caption{Merged results of dust activity and photometry of comet 3I/ATLAS by Lesedi and Skalnat\'{e} Pleso telescopes.}
    \label{tab:merged_table}
    \resizebox{\textwidth}{!}{%
    \begin{tabular}{c c c c c c c c c c c c c}
    \hline
    \makecell{Date\\(2025)} & \makecell{Telescope} & \makecell{$\rho^{*}$,\\km} & \makecell{$m_\mathrm{R}$,\\mag} & \makecell{$H_\mathrm{R}$,\\ mag} & \makecell{$Af(0^\circ)\rho$\\ ({\it g}), cm} & \makecell{$Af(0^\circ)\rho$\\ ({\it r}), cm} & \makecell{$Af(0^\circ)\rho$\\ ({\it i}), cm} & \makecell{$Af(0^\circ)\rho$\\ ({\it V}), cm} & \makecell{$Af(0^\circ)\rho$\\ ({\it R}), cm} & \makecell{$C_\mathrm{r}$,\\ km$^2$} & \makecell{$C_\mathrm{r}$,\\ km$^2$\ kg\,s$^{-1}$} & \makecell{$R_{c}$,\\ km }\\
    \hline
    Jul 2  & Lesedi         & --    & --             & --             & $276 \pm 12$ & $271 \pm 12$ & $283 \pm 14$ & --            & --            & 234 & $\sim (0.3 - 3.0)$ & <9 \\
    Jul 3  & Skalnat\'{e} Pleso & 10606 & 17.64$\pm$0.01 & 11.75$\pm$0.01 & --           & --           & --           & --            & $275 \pm 24$  & 333   & $\sim (0.3 - 2.9)$                & <10 \\
    Jul 5  & Skalnat\'{e} Pleso & 10418 & 17.47$\pm$0.01 & 11.65$\pm$0.01 & --           & --           & --           & $308 \pm 19$  & $315 \pm 24$  & 375    & $\sim (0.3 - 3.3)$                  & <11 \\
    Jul 7  & Lesedi         & --    & --             & --             & $329 \pm 15$ & $400 \pm 22$ & $407 \pm 22$ & --            & --            & 308 & $\sim (0.4 - 4.2)$ & <10 \\
    Jul 15 & Lesedi         & --    & --             & --             & $310 \pm 13$ & $384 \pm 13$ & --           & --            & --            & 237 & $\sim (0.3 - 3.7)$ & <9 \\
    Jul 20 & Skalnat\'{e} Pleso & 11640 & 17.33$\pm$0.01 & 12.01$\pm$0.01 & --           & --           & --           & $280 \pm 35$  & $275 \pm 15$  & 329    & $\sim (0.3 - 2.9)$              & <10 \\
    Jul 22 & Skalnat\'{e} Pleso & 11490 & 17.22$\pm$0.01 & 11.97$\pm$0.01 & --           & --           & --           & $267 \pm 23$  & $300 \pm 19$  & 355    & $\sim (0.3 - 3.1)$                 & <11 \\
    Jul 23 & Lesedi         & --    & --             & --             & $275 \pm 5$  & $347 \pm 4$  & --           & --            & --            & 431 & $\sim (0.5 - 6.9)$ & <11 \\
    Jul 23 & Skalnat\'{e} Pleso        & --    & --             & --             & -- & -- & --           & $270 \pm 27$             & $303 \pm 26$           & 359 & $\sim (0.3 - 3.1)$ & <11 \\
    \hline

    \textbf{Average} & -- & -- & -- &  -- & \textbf{298$\pm$11} & \textbf{351$\pm$13} & \textbf{345$\pm$18} & \textbf{281$\pm$26} & \textbf{294$\pm$22} & \textbf{329} & \textbf{$\sim (0.3 - 3.7)$} & \textbf{<10}\\
    \hline
\multicolumn{9}{l}{\footnotesize $^*$ $^*$ - the projected size of the photometric aperture at the distance of the comet.} \\    
    \end{tabular}%
    }
\end{table*}

The comet magnitudes in various filters on 02, 07, and 15 July 2025, measured within an aperture ranging from 10,300 to 11,600~km in radius centered on the optocentre, are listed in Table~\ref{tab:table_phot_combined}. Standard star magnitudes were taken from the APASS catalog \citep{henden2016vizier}, with an average photometric uncertainty of 0.02~mag.

Absolute magnitudes were computed as:

\begin{equation}\label{eq:H}
H_{f} = m_{f} - 5 \log_{10}(r_{h} \Delta) - \beta \alpha,
\end{equation}

where $r_{h}$ and $\Delta$ are the heliocentric and geocentric distances (in au), $\alpha$ is the phase angle, and $\beta = 0.04$~mag/deg is the assumed phase coefficient.

The dust production proxy $Af\rho$ was calculated following \citet{AHearn1984}:

\begin{equation}\label{eq:Afro_ph}
Af\rho = \frac{4 r_{h}^2 \Delta^2}{\rho} \times 10^{0.4 (m_{\odot} - m_c)},
\end{equation}

where $m_{\odot}$ and $m_c$ are the magnitudes of the Sun and the comet, and $\rho$ is the aperture radius. Solar magnitudes were taken from \citet{Willmer2018}. Phase corrections to $0^\circ$ phase angle were applied using the function from \citet{schleicher2011composition}. Results are provided in Tables~\ref{tab:table_phot_combined} and \ref{tab:merged_table}.

The scattering cross-section $C$ (km$^2$) was derived following \citet{jewitt2016fragmentation}:

\begin{equation}\label{eq:CS}
C = \frac{1.5 \times 10^{6}}{p_v} \times 10^{-0.4 H},
\end{equation}

where $p_v = 0.04$ is the geometric albedo. The corresponding radius is $R_c = \sqrt{C/\pi}$, representing an upper limit to the nucleus size.

Dust ejection velocity estimates $v_e$ follow \citet{jewitt1987surface} and \citet{hsieh2021physical}, yielding velocities between $\sim$0.012 and 1.2~m\,s$^{-1}$ for particle sizes from millimeters to meters.

The dust mass loss rate $\frac{dM}{dt}$ was calculated following \citet{fink2012calculation}:

\begin{equation}
\frac{dM}{dt} = \frac{4 \pi \rho_d v_e r_d A(0^\circ) f \rho}{3 p_v},
\end{equation}

assuming a bulk dust density $\rho_d = 1000$~kg\,m$^{-3}$ \citep{ivanova2016photometric} and $p_v = 0.04$. Results are reported in Table~\ref{tab:merged_table}.

\section{Determination of the spin state and lightcurve amplitude}

We compiled all observations obtained in the SDSS $r'$ band to derive the lightcurve of 3I/ATLAS. For some datasets, particularly those obtained during the first week of July, image stacking was employed to increase the signal-to-noise ratio (SNR) of the comet. All individual frames were visually inspected to identify and exclude those affected by stellar contamination, which was common due to the comet's motion toward the Galactic plane. To maximize the signal from the cometary nucleus while minimizing the contribution from the surrounding coma and background noise, photometric measurements were performed using an optimized circular aperture matched to the object's point spread function (PSF). Specifically, the aperture radius was set to encompass approximately one standard deviation ($1\sigma$) of the full width at half maximum (FWHM) of the comet's brightness profile in the stacked image. Photometric uncertainties were computed by combining photon noise, background noise, and scatter in the reference star magnitudes.

\begin{figure}
\centering
\includegraphics[angle=0,width=0.5\textwidth]{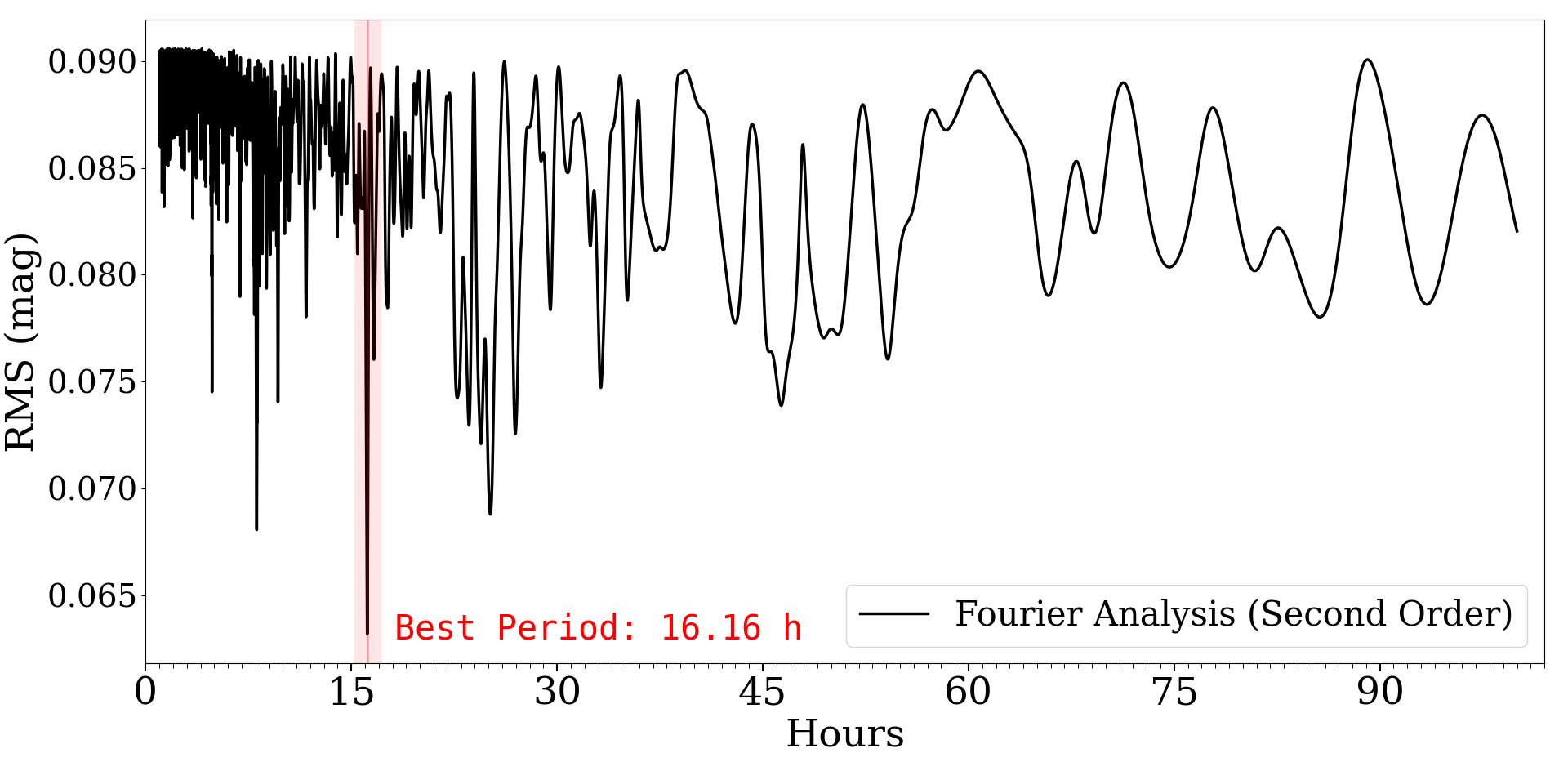}   
\caption{Periodogram obtained through a second-order Fourier analysis of the lightcurve data. The black curve represents the root mean square (RMS) residuals as a function of the tested rotation period, measured in hours. A clear minimum is observed at $16.16 \pm 0.01$ hours, indicated by the red shaded region.}
\label{fig:periodogram}
\end{figure}

The photometric measurements were reduced by applying light-time correction to the observation epochs, as well as distance corrections for the varying heliocentric and geocentric distances. A phase correction was applied using the H–G photometric system \citep{Bowell1989}, assuming a standard slope parameter of $G = 0.15$. To determine the best-fitting rotation period, we performed a Fourier analysis using a second-order harmonic function (see Fig~\ref{fig:periodogram}). 

In some lightcurve segments—particularly from F65 and Z23—the rotational modulation was not clearly detected, showing a relatively flat curve clustering near the average magnitude. This likely results from observing conditions (e.g., seeing, telescope resolution) and intrinsic comet activity, which enhances the coma and dilutes nucleus variability. The extensive Z23 dataset was probably acquired under such conditions, partially or entirely suppressing the rotational signature. Inclusion of these flat segments reduced the rotational contrast and introduced ambiguity in the Fourier analysis, likely contributing to alias solutions.

\end{appendix}
\end{document}